\documentclass[sigconf]{acmart}

\AtBeginDocument{%
  \providecommand\BibTeX{{%
    \normalfont B\kern-0.5em{\scshape i\kern-0.25em b}\kern-0.8em\TeX}}}

\setcopyright{acmcopyright}
\copyrightyear{2018}
\acmYear{2018}
\acmDOI{XXXXXXX.XXXXXXX}

\acmJournal{POMACS}
\acmVolume{37}
\acmNumber{4}
\acmArticle{111}
\acmMonth{8}

\makeatletter
\let\@authorsaddresses\@empty
\makeatother

\usepackage{pifont}
\usepackage{tabularx}

\usepackage{tabularray}
\usepackage{color}
\usepackage[normalem]{ulem}

\usepackage{csquotes}
\usepackage{float}

\usepackage{booktabs}
\usepackage{graphicx, subcaption}

\DeclareMathOperator{\coef}{coef}

\begin{document}

\title[What Are We Optimizing For?]{What Are We Optimizing For? A Human-centric Evaluation of Deep Learning-based Movie Recommenders}

\author{Ruixuan Sun}
\affiliation{%
 \institution{Grouplens Research, University of Minnesota}
 \streetaddress{5-244 Keller Hall, 200 Union Street SE}
 \city{Minneapolis}
 \state{Minnesota}
 \country{United States}}

\author{Xinyi Wu}
\affiliation{%
 \institution{
    Institute for Data, Systems, and Society (IDSS), Massachusetts Institute of Technology}
 \country{United States}}

\author{Avinash Akella}
\affiliation{%
 \institution{Grouplens Research, University of Minnesota}
 \streetaddress{5-244 Keller Hall, 200 Union Street SE}
 \city{Minneapolis}
 \state{Minnesota}
 \country{United States}}

\author{Ruoyan Kong}
\affiliation{%
 \institution{Grouplens Research, University of Minnesota}
 \streetaddress{5-244 Keller Hall, 200 Union Street SE}
 \city{Minneapolis}
 \state{Minnesota}
 \country{United States}}

\author{Bart Knijnenburg}
\affiliation{%
 \institution{School of Computing, Clemson University}
 \city{Clemson}
 \state{South Carolina}
 \country{United States}}

\author{Joseph A. Konstan}
\affiliation{%
 \institution{Grouplens Research, University of Minnesota}
 \streetaddress{5-244 Keller Hall, 200 Union Street SE}
 \city{Minneapolis}
 \state{Minnesota}
 \country{United States}}

\renewcommand{\shortauthors}{Sun, et al.}

\newcommand{\x}[1]{{\color{blue}[Xinyi: #1]}}

\begin{abstract}

In the past decade, deep learning (DL) models have gained prominence for their exceptional accuracy on benchmark datasets in recommender systems (RecSys). However, their evaluation has primarily relied on offline metrics, overlooking direct user perception and experience. To address this gap, we conduct a human-centric evaluation case study of four leading DL-RecSys models in the movie domain. We test how different DL-RecSys models perform in personalized recommendation generation by conducting survey study with 445 real active users. We find some DL-RecSys models to be superior in recommending novel and unexpected items and weaker in diversity, trustworthiness, transparency, accuracy, and overall user satisfaction compared to classic collaborative filtering (CF) methods. To further explain the reasons behind the underperformance, we apply a comprehensive path analysis. We discover that the lack of diversity and too much serendipity from DL models can negatively impact the consequent perceived transparency and personalization of recommendations. Such a path ultimately leads to lower summative user satisfaction. Qualitatively, we confirm with real user quotes that accuracy plus at least one other attribute is necessary to ensure a good user experience, while their demands for transparency and trust can not be neglected. Based on our findings, we discuss future human-centric DL-RecSys design and optimization strategies.  

\end{abstract}



\maketitle

\section{Introduction}

  For modern recommender systems (RecSys), deep learning (DL) models are commonly recognized as state-of-the-art (SOTA) solutions, usually credited to their high accuracy scores (e.g., RMSE, HR, recall, MRR, NDCG) on benchmark datasets \cite{recommender2024datasets}. Many DL models use user context such as profile and demographic information to generate personalized recommendations~\cite{villegas2018characterizing} and are optimized for beyond-accuracy measures. Yet, their quality has often been solely evaluated under standard offline metrics such as content coverage, novelty, diversity, and serendipity \cite{kaminskas2016diversity, ziarani2021deep}. 
    
    Before DL models became popular in the field of RecSys, researchers had evaluated RecSys algorithms beyond offline scores from users' perspective \cite{pu2012evaluating, knijnenburg2012explaining}. Some works design sets of metrics that cover user perception on recommendation performance, such as transparency and trust~\cite{pu2011user,sonboli2021fairness,nilashi2016recommendation}. Although these metrics reflect user perspectives better than offline ones, they overlook contextual factors like interaction history, profile data, recommendation preferences, and individual usage patterns. These elements are key to understanding long-term user behavior \cite{raza2019progress}.

    As a result, to leverage user contextual information into the RecSys evaluation process, some other works assessed user experience with human-centric frameworks taking personal characteristics into account on classic recommendation models such as matrix factorization \cite{knijnenburg2012explaining} and collaborative filtering \cite{nguyen2018user}. This work proved effective in building a relationship between user characteristics and perception or experience.  

    Since DL RecSys was not subject to substantial user-centric evaluation before, we design this study to comprehensively assess the performance of four types of DL models from a open-sourced SOTA leaderboard. We recruited 445 active real users from an online movie recommender and construct a questionnaire of 4 lists containing personalized top recommendations generated by different DL and baseline CF models. Our evaluation metrics are inspired by previous human-centric approaches \cite{knijnenburg2012explaining,nguyen2018user,pu2012evaluating} and focus on seven human-centric metrics including \emph{Novelty, Diversity, Serendipity, Accuracy, Trustoworthiness, Transparency}, and their overall \emph{Satisfaction} on the recommendations. Besides, we also designed a set of questions to elicit their preferred level of usage and recognized importance on each perception metric.
    
    To further analyze the latent relationship between user context and model-wise evaluation metrics, we also run path analysis by extracting and categorizing user contextual data into two major types of features: \emph{Observed usage history} and \emph{Expressed Preference Over Meta Recommender attributes}. With this study design, we aim to answer the following two major research questions:
    
        \begin{quote}
            \textbf{RQ1}: \emph{How do different DL-RecSys models perform compared with each other and classic CF methods, as evaluated by our human-centric framework?}
        
        \textbf{RQ2}: \emph{Which contextual factors impact users' perception or experience on DL-based recommendations?}
        \end{quote}

    \textbf{We summarize the contribution of our paper as follows}:
    \begin{itemize}
        \item We recruited 445 users to evaluate sets of recommendations from DL and classic CF algorithms, completing a survey about the recommendations and their overall preferences. We find that DL models win in recommending novel and unexpected items, but do not outperform classic methods regarding diversity, trustworthiness, transparency, accuracy, and general user satisfaction.
        

        \item We conduct a thorough path analysis of how each DL model affect different recommendation attributes. We find that low diversity and high serendipity in DL models can directly impacting user-perceived transparency of recommendations, followed by undermined trust and personalization, and finally contributing to lower satisfaction. Besides, users' average historic ratings and their recognized values of novelty and diversity in recommendation are also key contextual factors influencing their perceived serendipity, diversity, or novelty of movies.  
        
        \item Through qualitative analysis, we discover users' strong requests to have accuracy plus at least one other recommendation attributes, their demand for transparent recommendations, and their usage of accuracy and right-level of serendipity on building trust with the system. We also discuss some general design strategies for future human-centric DL-RecSys development. 

    \end{itemize}

    In the rest of the paper, we first discuss related work, then provide a high-level overview of our research methods and the four deep learning models. After that, we share the model-wise performance comparison based on statistical findings, contextual path analysis, and qualitative data analysis of real user evaluation data. Finally, we consolidate and discuss all findings and propose human-centric design implications for future DL-RecSys studies.
    
\section{Related Work}

\subsection{Evaluation of DL-RecSys}

    Deep neural networks are pervasively used in the RecSys domain for their nature to abstract user-item interaction patterns and effectively learn the representation of large amounts of input data \cite{zhang2019deep}. Depending on design choices, there exists a diverse set of architectural paradigms. Multilayer Perceptron (MLP)~\cite{gardner1998artificial} in recommendation systems separately encodes user and item features before combining them through fully connected layers to predict user preferences based on learned representations. Graph Neural Network (GNN) is one step further. It encodes real-world network structures, such as social networks or item relationships, with its graph structure and conduct link prediction for user-item recommendation tasks \cite{berg2017graph}. Recurrent Neural Network (RNN) is appropriate when it comes to sequential recommendation \cite{donkers2017sequential} for its ability to remember former computations in memory. Transformers work well in session-based recommendation tasks \cite{ludewig2021empirical} with its self-attention mechanism. When it comes to measuring the quality of recommendations, accuracy metrics such as error-based RMSE \cite{hodson2022root}, recall-based Hit Rate \cite{huang2021deep}, or ranking and precision-based NDCG \cite{huang2021deep} are treated as the major optimization targets.

    As~\citeauthor{FerrariDacrema2019AreWR} points out, systematic studies are needed for a fair evaluation of DL-RecSys models to truly assess the progress they bring to the field of recommender systems \cite{FerrariDacrema2019AreWR}. In prior literature, DL-RecSys model benchmarking was primarily offline,  centering around prediction accuracy~\cite{Zhu2022BARSTO, FerrariDacrema2019AreWR, Sun2022DaisyRec2B} and training time~\cite{Sun2022DaisyRec2B}. These works consistently found that DL-RecSys models are sometimes better than traditional methods regarding accuracy with well-tuned parameters. However, they cost much more to train, thus questioning their effectiveness and efficiency. Moreover, Netflix researchers conducted a case study of bag-of-items and sequential DL-RecSys models. They compared DL methods to classic approaches by measuring ranking improvements of all models in offline and online settings. Among many findings, they shared that while DL models are capable of generating good representations of time, range, or modalities, their generalization power can also amplify short-term objectives such as click prediction, while sacrifice long-term user satisfaction~\cite{steck2021deep}. 

\subsection{User Perspective in Recommender Systems}

    User experience has been a critical aspect in evaluating the success of recommendations ever since the field started. \citeauthor{konstan2012recommender} suggested that the only reliable way to measure the RecSys behavior in a natural context is through a long-term field experiment \cite{konstan2012recommender}. \citeauthor{munawar2020framework} identified that subjective recommender system aspects, such as perceived quality and effectiveness, can be significant factors in user satisfaction \cite{konstan2012recommender}. Similarly, \citeauthor{knijnenburg2015evaluating} and \citeauthor{pu2012evaluating} also proposed a user-centric evaluation framework to assess recommender systems with user experiments and statistical analysis \cite{knijnenburg2015evaluating,pu2012evaluating}. \citeauthor{kunkel2019let} evaluated differences of trustworthiness between personal and impersonal recommendations with real human explanations \cite{kunkel2019let}. In industry, RecSys practitioners mainly evaluated user values from their engagement \cite{kangas2021recommender,zou2019reinforcement,wu2017returning}, long-term satisfaction \cite{steck2021deep,garcia2018understanding}, and privacy \cite{spinelli2020youtube,bartlett2023analysing} as compliment to accuracy or monetization metrics.

\section{Research Methods}

    Our overall study design is split into two parts: 1) For data processing and model training, we generated personalized recommendations with four DL-RecSys models and two baseline collaborative filtering (CF) models for selected active users from an online movie recommender. 2) In phase two, we designed a user survey with top recommendation lists from each model for subjective evaluation, including both Likert-scale questions and free-form text input. Detailed user evaluation flow can be found in Fig. \ref{fig:study_diagram}.

\subsection{Deep Learning and Baseline Models}

    We select four distinct DL models from the SOTA leaderboard based on their accuracy performances with the major benchmark dataset MovieLens-1M (ML-1M)\footnote{ \url{https://paperswithcode.com/sota/collaborative-filtering-on-movielens-1m}}:

    \begin{itemize}
        \item \emph{\textbf{Neural Collaborative Filtering (NCF)}} \cite{he2017neural} is one of the early seminal works introducing DL methods to recommender systems. It employs deep learning for collaborative filtering (CF) by replacing the inner product with a neural network.
        \item \emph{\textbf{BERT4Rec}} \cite{sun2019bert4rec} is a deep bidirectional self-attention model to learn the representations for users' historical behavior sequences. One can trace the reasons for recommendation by checking the attention scores and finding the most important historical item in prediction.
        \item \emph{\textbf{SSE-PT}} \cite{wu2020sse} is a sequential-based personalized transformer. Similar to BERT4Rec, SSE-PT enjoys interpretability by allowing one to check the attention scores and find the most important historical item in prediction.
        \item \emph{\textbf{GLocal-K}} \cite{han2021glocal} focuses on feature extraction by generalizing and representing a high-dimensional sparse user-item matrix into a low-dimensional space. Efficiency in data sparsity is the key advantage of GLocal-K.
    \end{itemize}

    For each model, we take the code from the open-source repository and train it on the dataset we collect for real users (detailed in section \ref{section:user-train-data}). For fair comparison, we adopt the optimal hyperparameters for ML-1M dataset reported in their original papers. All code repository links and hyperparameters can be found in Appendix. We also summarize some claimed beyond-accuracy attributes from each model's original paper in Table \ref{tab:dl_model_feature_summary}.

    The two baseline models we choose are k-nearest neighbor user-based Collaborative Filtering (with min k = 2) \cite{ekstrand2011collaborative} and funk SVD \cite{funk2006} (with factors=10 and epochs=20) models \footnote{The two CF models can be found in the Surprise library at \url{https://surpriselib.com/}.}

    \begin{table}
    \resizebox{\columnwidth}{!}{%
    \begin{tabular}{@{}lcccc@{}}
    \toprule
    \textbf{Feature\textbackslash Model} & \multicolumn{1}{l}{NCF} & \multicolumn{1}{l}{SSE-PT} & \multicolumn{1}{l}{BERT4Rec} & \multicolumn{1}{l}{GLocal-K} \\ \midrule
    \textit{Personalization}  & \ding{52} & \ding{52} & \ding{52} & - \\
    \textit{Timeliness*}      & \ding{52} & \ding{52} & \ding{52} & - \\
    \textit{Interpretability} & - & \ding{52} & \ding{52} & - \\
    \textit{Serendipity}      & - & - & - & - \\
    \textit{Diversity}        & - & - & - & - \\ 
    \textit{Efficiency}       & - & - & - & \ding{52} \\
    \textit{Cold-start}       & - & - & - & \ding{52} \\ 
    \bottomrule
    \end{tabular}%
    }
    \caption{Attribute comparison of DL-RecSys models. A check mark means the original paper of this model claimed to have achieved the attribute. *For timeliness, as it was implicit in the evaluation task, we referred to the good performance of sequential prediction mentioned in the paper.}
    \label{tab:dl_model_feature_summary}
    \end{table}

\subsection{Users and Training Data} \label{section:user-train-data}

    The participants recruited for this study were from MovieLens (\url{https://movielens.org/}), an academic-running online movie recommender system with thousands of active users~\footnote{We appreciate the support provided by the MovieLens team for this study.}. Due to the strict user privacy policy of the website, we select participants mainly based on their activity level (logged in over 12 times and rated over 20 movies) in the year 2022. As the base training dataset, we then collect those active users' ratings in the three calendar years before the experiment (from 2020-01-01 until 2022-12-31). To ensure the minimum popularity of movies, we also filter out those with less than 20 user ratings. The dataset contains 3,537 users, 7,462 movies, and 983,376 ratings. With the new real user dataset, we generate a personalized list containing the top-recommended movies for each user for the four DL models and two baseline CF models. To avoid making the questionnaire too long to exceed the general user's attention span, we only randomly assign 3 DL-generated recommendation lists and 1 CF list to each user.

    We sent out surveys via email to qualified 3,537 active users in 7 batches between April to June 2023, with one week between consecutive batches. In the email, we emphasized that filling in this survey was voluntary with no incentives, and users had the right to exit at any time during their participation. We then collected user responses after two weeks of the survey distribution. Overall, 3,172 out of 3,537 surveys were successfully delivered to users' email inboxes, and 445 of them replied, making the final response rate as \textbf{14.03\%}. Our Institutional Review Board (IRB) evaluated the experiment design and determined it to be a non-human subject study. 

\begin{figure*}[t]
        \centering
        \includegraphics[width=1.0\textwidth]{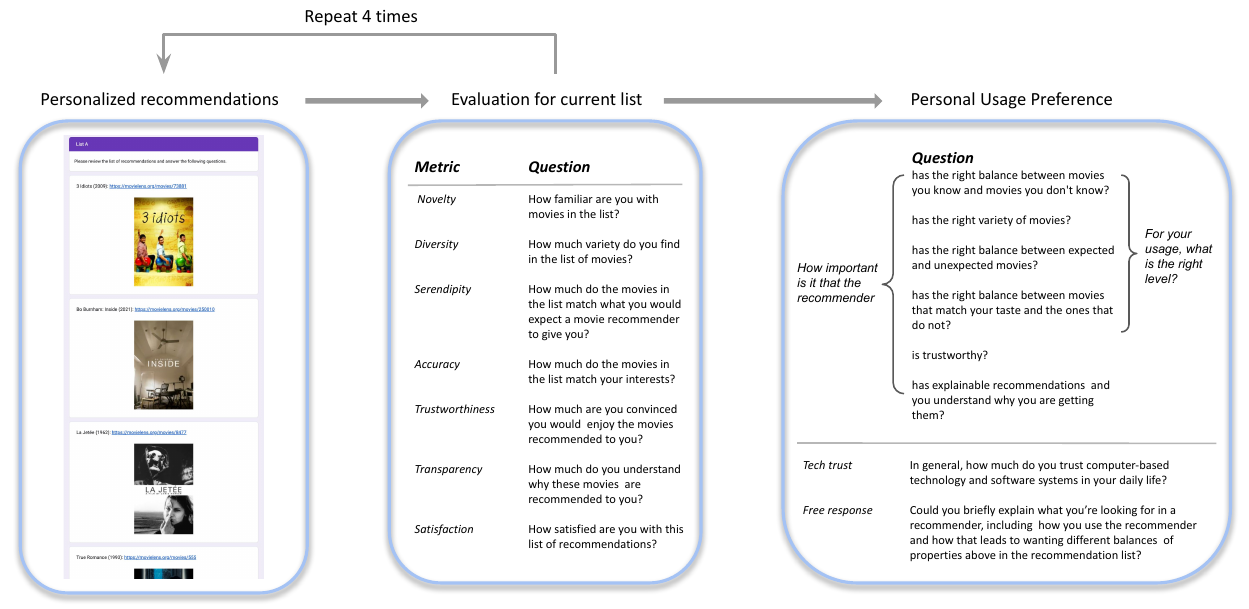}
        \caption{User evaluation flow and survey questions.}
        \label{fig:study_diagram}
    \end{figure*}

\subsection{Survey Design}

    Every user gets a Google form survey of 6 pages. The first page describes the purpose of this study ("test different personalization models"). After that, users proceed to 4 pages of recommendations that each contains a list of the top 12 movies (we ran pilot tests and considered that to be an adequate number for users to consume and make judgments) generated by one of the three DL models or one CF model. Each recommendation item contains the movie name, release year, and detailed MovieLens link that users can visit to check for details. After reviewing each page of recommendations, users are asked to fill in 7 human-centric questions outlined in Fig. \ref{fig:study_diagram}. Our design and phrasing of questions was inspired by a series of previous evaluation and user study works \cite{cremonesi2011looking,pu2011user,said2013user,avazpour2014dimensions,ekstrand2014user,knijnenburg2015evaluating,zolaktaf2018bridging} that include both item attributes (e.g. how novel or diverse a list of recommendations look like) and more subjective human perception (e.g. how transparent or trustworthy the results are), along with a summative user experience indicator (i.e. satisfaction). 
    
    The last page of the survey asks how important users consider each human-centric metric and the ideal level of each metric for them based on individual preference. We do not ask about the level for \emph{Trustworthiness} and \emph{Transparency}, since previous research suggested higher values in those two metrics yielded better recommendation results and user engagement \cite{victor2011trust,mohammadi2019trust,sonboli2021fairness,lepri2018fair,schnabel2020impact}. To best of our knowledge, we are the first study asking both the importance and ideal level from users to understand their separate impacts on user perception and satisfaction of recommendation qualities. At the end, we also provide an optional text field for users to share their expected recommendation attributes. 
    
     All questions are designed on a 5-point Likert scale except for free text responses. Specifically, the first question on the third list of recommendations is designed as an attention check. It is a reverse-scaled \cite{abbey2017attention} 5-point Likert question asking the same \emph{Satisfaction} question in a different phrase: \emph{"How much do you like the list of recommendations?"} In the analysis stage, we compare the reversed response between this check question and the actual satisfaction question on the third list of recommendations from all users with a Mann-Whitney-Wilcoxon test \cite{de2010five} and confirm they are not statistically different ($p=0.664$), meaning that general user attention during the survey is high.

\subsection{Path Analysis} \label{sec:path-analysis}

  \begin{table}[]
    \resizebox{\columnwidth}{!}{%
    \begin{tabular}{ll}
    \toprule
    \textit{\textbf{Feature}} & \textit{\textbf{Description}}                                 \\ \midrule
    \textit{tenure}           & Number of months since user registered on MovieLens            \\
    \\
    \textit{avg\_rating}      & Mean of one user's all historic ratings                                   \\
    \\
    \textit{genres\_cnt} &
      \begin{tabular}[c]{@{}l@{}} Number of genres (out of 20) that make up top 50\% \\ of the cumulative sum of all movies a user rated\end{tabular} \\
      \\
    \textit{login\_freq}      & Count of logins per month since their first login to the site \\
    \\
    \textit{rating\_freq}     & Count of ratings added per login session                      \\
    \\
    \textit{wishlist\_freq}   & Count of movies added to wishlist per login session           \\ \bottomrule
    \end{tabular}%
    }
    \caption{Users' observed usage history with the system.}
    \label{tab::user_observed_usage_history}
    \end{table}
    
Based on prior work about relevant user contextual features \cite{zhao2018explicit}, we categorize accessible user features from the MovieLens database into two types: \emph{1) Observed Usage History, 2) Expressed Preference Over Meta Recommended Attributes}. The first type of feature is defined in Table \ref{tab::user_observed_usage_history}, and the questions for getting the second type of feature can be found in Fig. \ref{fig:study_diagram}. Following \citeauthor{knijnenburg2012explaining}'s user-centric evaluation framework of recommender systems \cite{knijnenburg2012explaining}, we further bucket our data into five aspects: Objective System Aspects (OSA), Subjective System Aspects (SSA), Experience (EXP), Personal Characteristics (PC), and Situational Characteristics (SC), as shown in Table \ref{tab:var_mapping}. We then run path analysis \cite{lleras2005path} with MPlus statistic tool \cite{muthen2017mplus} based on the interaction flow suggested by \citeauthor{knijnenburg2012explaining} to quantify the impact within and between each contextual feature clusters to user's perception and experience of the seven human metrics \cite{knijnenburg2012explaining}. To understand individual model effect further, we create binary dummy variables of five different models (four DL models plus user-based CF) against the SVD baseline (as it surpassed others in most metrics in Table \ref{tab::user_eval_stats_table}). We start the model construction by building saturated paths \cite{knijnenburg2015evaluating} within SSA factors and expand to other data types. During each improvement iteration, we apply backward elimination and only keep the variables demonstrating significant direct effects until the model converges: $\beta_{\coef} >= .10$ or $\beta_{\coef} <= -.10$ and $p <= .05$. Finally, we use "categorical" option to account for the Likert-scale nature of our data and apply "STDY" for standardizing the output.

    \begin{table}[]
    \centering
    \resizebox{\columnwidth}{!}{%
    \begin{tabular}{@{}cll@{}}
    \toprule
    \textbf{Feature Group} &
      \multicolumn{1}{c}{\textit{\textbf{Variable}}} &
      \multicolumn{1}{c}{\textbf{Direct Impact Factors}} \\ \midrule
    OSA &
      \textit{\begin{tabular}[c]{@{}l@{}}is\_\{NCF, BERT4Rec, \\      GLocal-K, SSE-PT\}\end{tabular}} &
      -- \\ \midrule
    SSA & \textit{\begin{tabular}[c]{@{}l@{}}Novelty, Diversity, \\ Serendipity, Accuracy,\\ Trustworthiness, Transparency\end{tabular}} & OSA, SC, PC \\ \midrule
    EXP &
      \textit{Satisfaction} &
      SSA, SC, PC \\ \midrule
    PC &
      \textit{\begin{tabular}[c]{@{}l@{}}Observed system usage, \\ Recognized importance, \\ Tech Trust\end{tabular}} &
      -- \\ \midrule
    SC &
      \textit{Preferred Usage Level} &
      -- \\ \bottomrule
    \end{tabular}%
    }
    \caption{Mapping and direct effects between variables to user-centric evaluation framework feature groups}
    \label{tab:var_mapping}
    \end{table}

\section{Results} \label{sec:results}

\subsection{Individual Model Performance}

\begin{table*}
\resizebox{2\columnwidth}{!}{%
\begin{tabular}{@{}lllllll@{}}
\toprule
\textit{\textbf{}} & \textbf{NCF(n=337)} & \textbf{SSE-PT(n=317)} & \textbf{BERT4Rec(n=340)} & \textbf{GLocal-K(n=334)} & \textbf{SVD(n=208)} & \textbf{UU-CF(n=237)} \\ \midrule
\textit{Novelty}         & 3.104±1.104(***) & 2.535±0.955(***)     & 2.974±1.119(***) & \textbf{3.611±1.010} & 2.755±1.143(***)     & 2.827±1.146(***)     \\
\textit{Diversity}       & 3.656±0.985(***) & 3.972±0.836(*)          & 3.657±0.983(***) & 3.428±0.992(***)     & 3.957±0.934          & \textbf{4.004±0.916} \\
\textit{Serendipity}     & 2.504±0.897(***) & \textbf{3.252±1.063} & 2.674±1.010(***) & 2.361±0.853(***)     & 2.611±0.899(***)     & 2.574±1.000(***)     \\
\textit{Trustworthiness} & 3.473±1.042      & 2.877±1.035(***)     & 3.095±1.012(**) & 3.144±0.996(**)     & \textbf{3.495±0.863} & 3.228±1.024(*)       \\
\textit{Transprency}     & 3.516±1.066      & 2.397±1.212(***)     & 2.926±1.132(***) & 3.197±0.980(**)     & \textbf{3.587±0.928} & 3.245±1.065(*)     \\
\textit{Accuracy}        & 3.214±1.201      & 2.817±0.986(***)     & 2.808±1.047(***) & 2.875±1.125(***)     & \textbf{3.284±1.228} & 3.013±1.152(*)       \\
\textit{Satisfaction}    & 3.458±1.030      & 3.132±1.084(**)     & 3.118±0.974(*) & 3.144±1.003(***)     & \textbf{3.500±0.896} & 3.160±1.017(**)     \\ \bottomrule
\end{tabular}%
}
\caption{Means and STDs for DL model recommendation evaluation. For each metric, the highest value is highlighted in bold, and a path model from MPlus is applied between each of the rest of the model's data to the one with the highest value for statistical significance assessment. For the last 4 metrics where NCF shows no significant difference from SVD, we tested with a separate path model and confirmed NCF has a significant difference from the other three DL models. Asterisk (*) indicates p-val between models:  * for $p$ < .05, ** for $p$ < .01 and *** for $p$ < .001.}
\label{tab::user_eval_stats_table}
\end{table*}

 We report the mean and standard deviation (STD) along with pairwise statistical significant difference of all model-wise user evaluation questions (see Fig. \ref{fig:study_diagram}) in Table \ref{tab::user_eval_stats_table}. With that, we answer \textbf{RQ1}:
 
\begin{quote}
\textbf{RQ1}:  \emph{How do different DL-RecSys models perform compared with each other and classic CF methods, as evaluated by our human-centric framework?}
\end{quote}
 
 In Table \ref{tab::user_eval_stats_table}, we observe that GLocal-K is the top performer in \emph{Novelty}, while SSE-PT wins in \emph{Serendipity}. Without CF baselines, NCF perfroms the best in terms of delivering diverse, trustworthy, transparent, accurate recommendations, while SSE-PT and BERT4Rec perform worst on \emph{Trustworthiness}, \emph{Transparency}, \emph{Accuracy}, and \emph{Satisfaction}. With this preliminary comparison, we see that DL models might outperform in recommending new or unexpected items, but are worse at producing a diverse set of items, gaining user trust or transparency, and most importantly, matching user personalized interest and achieving high satisfaction.
 

\subsection{User Perception Path}

 Based on individual model performance, we are curious to understand what latent variables contribute to user perception and experience. Since metrics such as trust, accuracy, and transparency can only be assessed by movies which users have already seen, we hypothesize that users' perception of them should be built with more directly accessible characteristics like novelty and diversity. Subsequently, all recommendation quality attributes can be strong factors influencing the holistic measure of user general satisfaction. Moreover, user behaviors and preferences, such as their interactive patterns with the system and their subjective preference for different RecSys attributes, can also be possible contributors to their final judgment on how good the recommendation quality is.
 
\begin{figure*}[]
        \centering
        \includegraphics[width=2\columnwidth]{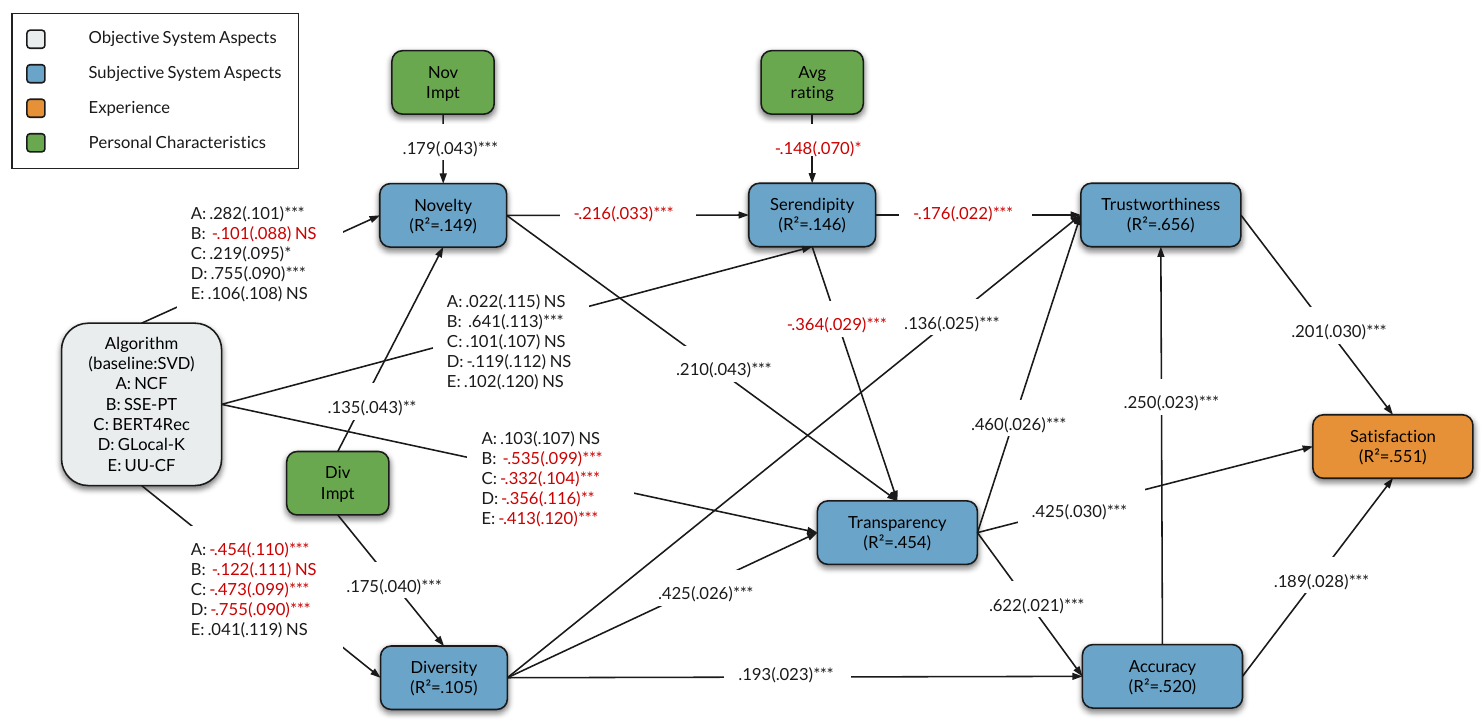}
        \caption{Path analysis on data we have from this study. The model shows how user contextual factors and different model perception metrics can influence each other and overall user satisfaction. Each arrow indicates a direct effect between one variable to another, with the $\beta_{\coef}$ and standard error associated on the line. Asterisk (*) indicates effect p-val:  * for $p$ < .05, ** for $p$ < .01 and *** for $p$ < .001.}
        \label{fig:sem}
\end{figure*}

 Based on our hypothesis, we construct a path model with methods detailed in Section \ref{sec:path-analysis}. As illustrated in Fig. \ref{fig:sem}, it has a good statistical fit, with $\chi^2(100)=110.143, DoF=40, p < 0.001, CFI = 0.987, TLI = 0.975, RMSEA = 0.039, 90\% - CI: [0.030, 0.048]$. We then answer \textbf{RQ2} with the path analysis:

\begin{quote}
    \textbf{RQ2}: \emph{Which contextual factors impact users' perception or experience on DL-based recommendations?}
\end{quote}

    \begin{figure}[]
        \centering
        \includegraphics[width=1.0\columnwidth]{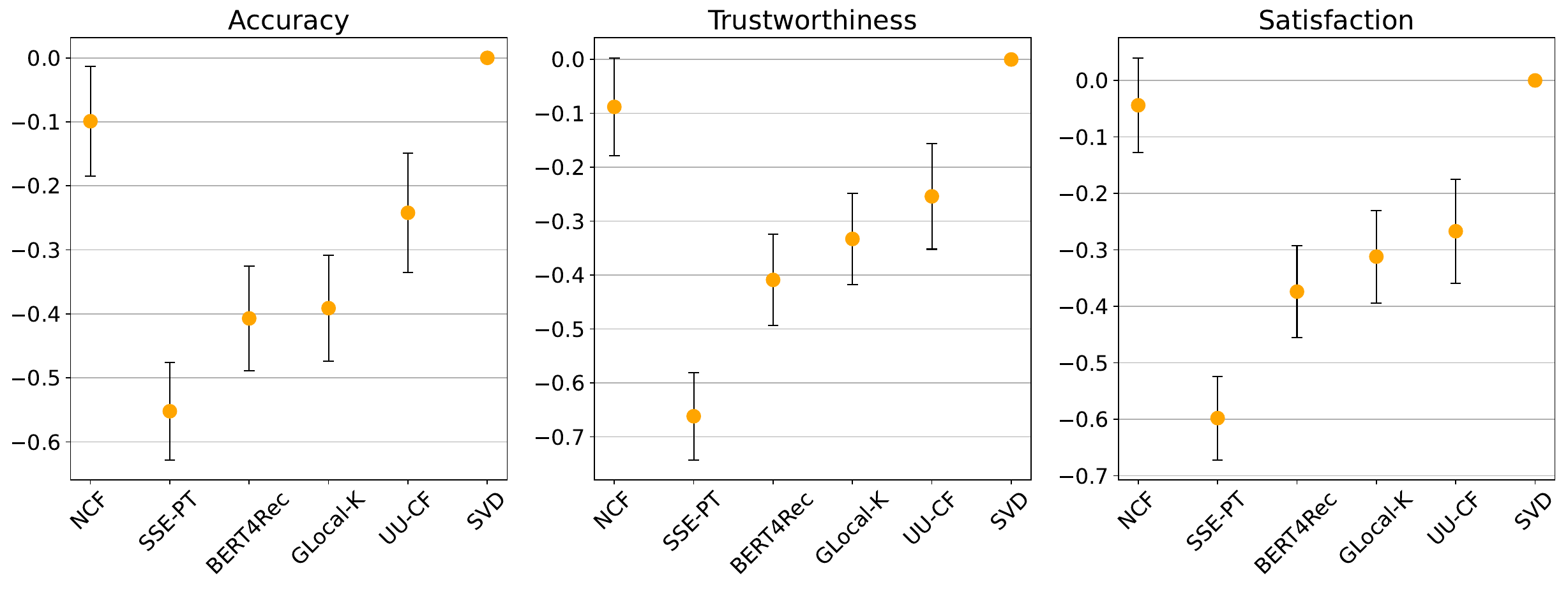}
        \caption{Marginal effects (both direct and indirect) of each model to the downstream variables in the path. Error bars indicate standard errors.}
        \label{fig:path_marginal_models}
    \end{figure}

    \begin{table}[]
        \resizebox{\columnwidth}{!}{%
        \begin{tabular}{@{}lccc@{}}
        \toprule
        \textit{\textbf{}} & \textbf{Accuracy} & \textbf{Trustworthiness} & \textbf{Satisfaction} \\ \midrule
        \textit{avg\_rating} & .034(.016) & .059(.028) & .041(.020) \\
        \textit{nov\_impt}   & .032(.009) & .039(.011) & .036(.010) \\
        \textit{div\_impt}   & .104(.025) & .107(.025) & .090(.022) \\ \bottomrule
        \end{tabular}%
        }
        \caption{Marginal effects (both direct and indirect) of user contextual factors to downstream variables in the path. Each cell includes the effect size and standard error.}
        \label{tab:path_marginal_context}
    \end{table}

Starting from user satisfaction, over half ($R^2=0.551$) of its contribution is from the paths connecting to three user perceived attributes: \emph{Accuracy, Trustworthiness}, and \emph{Transparency}. Within the three variables, \emph{Trustworthiness} can be directly contributed by \emph{Accuracy, Transparency}, and \emph{Diversity}, while \emph{Serendipity} undermine trust. Also, \emph{Accuracy} are majorly contributed by \emph{Transparency} and \emph{Diversity}. In the middle layer, we observe \emph{Novelty} and \emph{Diversity} are the two major positive contributors of \emph{Transparency}, while \emph{Serendipity} served as the negative one. We also see that all DL models other than NCF are negatively contributing to \emph{Transparency}, compared to the baseline SVD. As for the most direct perception metrics \emph{Novelty} and \emph{Diversity}, we confirmed what we found out earlier in Table \ref{tab::user_eval_stats_table} -- most DL models are capable of recommending novel items, but failed in generating a diverse set, compared to baseline SVD. Finally, when we digest the path as a whole, we see a clear track of why most DL models failed in providing satisfactory recommendations to users: the failure of generating diverse or transparent movies could impact the following accuracy or trustworthiness attributes, which can directly undermine final satisfaction. On the other hand, generating too much serendipitous recommendations might confuse users and weaken transparency and trust, finally causing worse user experience.



Apart from the algorithm factor, three user-native contextual variables can also influence how users perceive recommendations. Under all four DL models, users' recognized importance of novelty can positively impact how they perceive novelty from recommendations, while their recognized importance of diversity is positively related to both perceived recommendation novelty and diversity. Their average rating is observed to have negative effects on their perceived serendipity, indicating the lower their historic ratings are, the more unexpected they would find the recommendations generated from DL models.

To conclude the path analysis, we display the marginal effects of individual DL models and user contextual factors on the downstream variables \emph{Accuracy, Trustworthiness}, and \emph{Satisfaction} in Fig. \ref{fig:path_marginal_models} and Table \ref{tab:path_marginal_context}. The results suggest that recommendations from NCF, SSE-PT, BERT4Rec, GLocal-K, and UU-CF can marginally lead to lower appreciation in all three downstream attributes compared to the SVD baseline, confirming our findings from Table \ref{tab::user_eval_stats_table}. Besides, when user-recognized importance of diversity, novelty, and average ratings are higher, we also observe slightly higher total positive effects on \emph{Accuracy, Trustworthiness}, and \emph{Satisfaction}.

\subsection{Qualitative Analysis}

In the last section of real user evaluation, we look into users' commonly requested recommender properties and the balance between them. In total, we have 294 users who shared their preferences in the free response. Then, three researchers followed the grounded theory method (GMT) \cite{strauss1994grounded} to conduct open coding, inductive thematic analyses \cite{braun2012thematic}, and affinity map building of different clusters on codes with similar meanings. The process iterated until all inputs were clustered, and no ambiguity or disagreement emerged. Selected user quotes and their mapping recommender properties are displayed in Table \ref{tab::example_qual_input}.

    \begin{table*}[]
    \resizebox{\textwidth}{!}{%
    \begin{tabular}{@{}lll@{}}
    \toprule
    \textit{\textbf{Participant}} & \textit{\textbf{Property}}                                                                  & \textit{\textbf{Input}}                                                                                                                                                                                                                                                                                                                                                                  \\ \midrule
    \textit{P6}                   & \textit{\begin{tabular}[c]{@{}l@{}}Trustworthiness,\\ Diversity\end{tabular}}               & \begin{tabular}[c]{@{}l@{}}"I want a recommender that I can rely on to pick movies to watch. I want to see it recommend a variety of different \\ genres and styles that introduce me  to new movies I wouldn't have come across on my own but that I enjoy or can \\ learn from or otherwise appreciate."\end{tabular}                                                                  \\ \addlinespace
    \textit{P18}                  & \textit{Diversity}                                                                          & \begin{tabular}[c]{@{}l@{}}"Usually I've different moods at that situation and not that many movies will be the fit, so I guess the best would be \\ getting a good variety of movies that are really not similar between them"\end{tabular}                                                                                                                                             \\ \addlinespace
    \textit{P34}                  & \textit{Transparency}                                                                       & \begin{tabular}[c]{@{}l@{}}"If I am shown something that appears to be very different for me, I'm interested to understand why the movie was \\ selected for me - that can help me decide if I agree and want to watch it."\end{tabular}                                                                                                                                                 \\ \addlinespace
    \textit{P43}                  & \textit{Serendipity}                                                                        & \begin{tabular}[c]{@{}l@{}}"A black box that spits out movies I'll enjoy. When I don't have a good movie in mind, I sort the MovieLens list by \\ predicted rating, and pick one of the top ones that looks interesting"\end{tabular}                                                                                                                                                    \\ \addlinespace
    \textit{P68}                  & \textit{Accuracy}                                                                           & \begin{tabular}[c]{@{}l@{}}"To recommend movies for me to watch. I don't care how it decides this. I want to watch great stories with \\ good actors that are well-directed."\end{tabular}                                                                                                                                                                                               \\ \addlinespace
    \textit{P70}                  & \textit{Novelty}                                                                            & "I'm mostly looking to find things I haven't heard of or somehow missed."                                                                                                                                                                                                                                                                                                                \\ \addlinespace
    \textit{P75}                  & \textit{\begin{tabular}[c]{@{}l@{}}Novelty,\\ Accuracy\end{tabular}}                        & \begin{tabular}[c]{@{}l@{}}"I'd like the recommender to give me good list of movies that might spark interest in new genres by introducing me \\ to them with films that are somewhat familiar and somewhat new to me."\end{tabular}                                                                                                                                                     \\ \addlinespace
    \textit{P96}                  & \textit{Trustworthiness}                                                                    & \begin{tabular}[c]{@{}l@{}}"I use the recommenders as double-check devices: I find a movie; if I find it interesting I check the rating given by \\ the recommender. If it's good, I watch it. But I wouldn't trust the recommender on 'blind' recommendations."\end{tabular}                                                                                                            \\ \addlinespace
    \textit{P146}                 & \textit{\begin{tabular}[c]{@{}l@{}}Accuracy, \\ Trustworthiness, \\ Diversity\end{tabular}} & \begin{tabular}[c]{@{}l@{}}"I like to see movies recommended to me which I have already seen. This mostly just shows me that the algorithm \\ is in track and I can trust the movies it’s shown me that I don’t know. It would also be good to see a different set of \\ recommendations each time I visited the website."\end{tabular}                                                  \\ \addlinespace
    \textit{P186}                 & \textit{\begin{tabular}[c]{@{}l@{}}Transparency, \\ Accuracy, \\ Diversity\end{tabular}}    & \begin{tabular}[c]{@{}l@{}}"I'm looking for something to be able to make sense of the reasons why I like the movies that I like, and to not be afraid \\ of recommending me strange niche movies. But at the same time, if possible, to not recommend me things that \\ I clearly do not watch, while keeping variety."\end{tabular}   
    \\ \addlinespace
    \textit{P86}                  & \textit{\begin{tabular}[c]{@{}l@{}}Gap between \\ AI and human\end{tabular}}                                                                    & \begin{tabular}[c]{@{}l@{}}"I'm expecting a recommender to find movies that have good ratings from other users who's ratings are similar to mine. \\ I don't want recommendations based on information about the movie. Maybe someday AI will be capable enough to \\ watch movies and guess what I like about them, but till then only other users have that information."\end{tabular} 
    \\ \bottomrule
    \end{tabular}%
    }
    \caption{Examples of user input about important properties of recommendations.}
    \label{tab::example_qual_input}
    \end{table*}

  While many users only mentioned about their preference for accurate recommendations (N=52), we also see a great portion of others requesting the following three types of RecSys properties: 1) Urge for a good balance between accuracy with other metric(s), 2) Use of specific properties to build trust with the recommender, and 3) Demand for recommendation transparency.

  \textit{\textbf{Accuracy + X.}}\quad The largest theme under this cluster is about the balance between accuracy and novelty (N=54), like P28 shared: \emph{"I use it primarily to find out about movies I hadn't considered that closely match the kinds of movies I like."} The next is surrounding accuracy with diversity (N=12). For example, P39 said: "\emph{"I want the recommender to find movies I'm not familiar with that it thinks I will like. I want broad recommendations across lots of genres, and time (both old and new movies)."} The third one blends accuracy with serendipity (N=10) -- P234: \emph{"I want a recommender that would challenge my tastes without offending me."}

 \textit{\textbf{Trust builder.}}\quad Another big theme users discuss is to gain well-grounded recommendations (N=21). As P290 shares, \emph{"I want to get reliable recommendations of movies that I wouldn't have come across otherwise."} One step further, many users also mention other properties they relate to building trust with the recommender, such as accuracy (N=10), \emph{"I want the recommender to be adapted to my tastes so I can have a big level of confidence that I will enjoy the movies listed (P46)"}. P275 claim that their trust is built upon serendipity, \emph{"I would even go as far to say I 'trust' or enjoy MovieLens specifically because I can't tell where the recommendation came from. I think when it's traceable that's what feels fake or mechanical. "} Nevertheless, some people insist on putting their confidence more on real humans, rather than AI-powered algorithms (N=3), like P79 points out, \emph{"I use a recommender only as a second or third source. Reading and word of mouth are primary, and the recommender can effectively move things on an informal list of things to watch relative to their position based on the first couple of sources."}

   \textit{\textbf{Demand for transparency.}} \quad The third cluster we identify for preferred recommendation value is transparency (N=27). Some users appreciate more explanation about the recommendation, like P286 says, \emph{"...I would also love to see just a tiny bit more info on the films themselves, most importantly the writer(s). That would be a good tool to watch films from certain people...It would be cool if users had a better idea of what was going on under the hood as far as the recommendations go."} Aligned with that demand, more context of recommendation is also appreciated. For example, P132 suggests that they would consider recommendations around movies they have watched before to approach transparency: \emph{"Pointing out why it was licked would help me select a movie to be watched. Amazon Prime's recommendations on a specific movie I watched are helpful for example."} Some more proactive users request control over what can be generated, like P9 shares: \emph{"I want to have control over recommendations. Sometimes I am into fresh thrillers and therefore only want to see thrillers released after a certain year. Sometimes I would like to explore different genres of movies and therefore want to see very diverse recommendations."}

\section{Discussion}

    In this section, we consolidate our findings and discuss the state of DL-RecSys models from a human-centered perspective as well as opportunities for future design.

    \subsection{Current State of DL-RecSys models}

    While DL-RecSys models perform well in offline metrics, in our study, we find that they are not as good when evaluated with human-centric metrics. Compared to baseline CF models, only NCF performs relatively equal in terms of generating transparent, accurate, and trustworthy recommendations. It also has the highest performance in terms of general user satisfaction among all DL models. Since the other three DL models we assessed did not get close to CF in most metrics other than novelty and serendipity, there are still a lot of work to do with DL-RecSys user-centric improvement.

    In particular, though some DL models have superior performance in terms of serendipity, prior work by \citeauthor{kotkov2018investigating} suggested that higher serendipity in movies did not necessarily lead to higher user satisfaction \cite{kotkov2018investigating}. In our case, the high serendipity performance on SSE-PT also did not produce corresponding high satisfaction in that particular model, and the path analysis indicates a negative impact from perceived serendipity to user trust. For future DL-RecSys work focusing on serendipitous recommendation, it will be helpful to test carefully with real users to explore the right balance of serendipity, instead of blindly optimizing for a higher value.

    Overall, it is challenging to give specific optimization plan since different DL models have subtle and complex training logic, but by looking into some specific user cases, we summarize some generalized strategies researcher can consider for future development. One major demand from users asking for more transparency in algorithms can benefit from generating item descriptions with more personalized context, like P132 shares: \emph{"I expect the recommender to recommend me movies that I haven't seen but would enjoy. Pointing out why it was licked would help me select a movie to be watched"}. Thus, incorporating personalized recommendation explanations with new technique such as large language models \cite{silva2024leveraging,acharya2023llm} can be promising future direction to explore. In addition, trustworthiness can be built with more than accuracy in recommendations, as P6 and P186 shared in Table \ref{tab::example_qual_input}. We suggest future practitioners also look into improving diversity and transparency aspects to ensure a higher user trust.

\subsection{DL-RecSys Attribute Path}

    What the path analysis on different RecSys properties and user contextual factors implies can be split into two folds. When we look at users' observed usage history, we see that how long they have been in the system, or how often they visit and interact with the system, are not significant contributors to their perception of the recommendation quality or general satisfaction. Instead, their average historic rating resulting from their personal rating scale is one key factor influencing their perceived serendipity of recommendations. Besides, users' recognition of the importance of novelty and diversity positively correlates with their perceived novelty or diversity of recommendations. However, their ideal level of usage do not show significant effects with any recommendation attributes. Based on the findings, we suggest that future practitioners in DL-RecSys to gather more subjective user data, such as their rating dispositions \cite{sun2023less} and personality traits \cite{nguyen2018user}, to help DL models better learn target users' behavior and preference in the training phase.  
    
    On the other hand, recommendations generated from DL-RecSys models lack diversity in general, probably due to overfitting users' historical training data, which can compromise the downstream user-perceived transparency, trustworthiness, and accuracy of recommendations. For future optimization, incorporating an extra diversity balance mechanism in the training process such as diversity-aware re-ranking \cite{wang2020personalized,abdollahpouri2019managing,liu2022neural}, or providing diversity control UI to end users \cite{sun2023interactive,jin2020effects,harambam2019designing} can be beneficial for improving accuracy and summative user satisfaction.

\subsection{Clarification and Limitations}

    We want to reiterate and further clarify that we did not deliberately tune the hyperparameters of chosen DL models to optimize for offline metrics, which is not the focus of our study. Instead, we use the optimal values reported by the authors in the corresponding paper or repository. As \citeauthor{shehzad2023everyone} pointed out, careful tuning of DL models can usually make specific models outperform non-optimized baselines \cite{shehzad2023everyone}. However, since the main focus of this study is to reveal the traditionally ignored human-centric metrics that were not contained in the original measurement suits, we believe our findings carry significant meanings in terms of future optimization directions.

    Admittedly, we recognize some limitations of this study, including 1) We only chose DL-based models based on their performance of movie recommendation and collected real user feedback in the movie domain, without further expansion to other application fields; 2) We only tested personalized recommendations with active users, and did not generalize how those DL-RecSys models perform with cold-start users; 3) Due to reproducibility and time constraints, we did not evaluate other DL-RecSys models other than the mentioned four; 4) Our construct of the path model is only based on single-item survey responses instead of a more thorough multi-item CFA measurement; 5) Our survey design was static instead of a more interactive and dynamic UI. We believe all the above points can be interesting future works. For example, researchers can conduct cross-domain studies, select both new and old users as test subjects, and design larger-scaled A/B testing with multiple survey questions to assess each dimension of measurement.  

\section{Conclusion}

In this study, we investigate how four SOTA DL-RecSys models perform under multi-dimensional human-centric evaluation with a movie recommendation case study. By producing personalized top recommendation lists and evaluating them with real user feedback, we find that sequential and kernel-based DL-RecSys models are superior in recommending novel and serendipitous items while underperforming classic CF models in user-perceived diversity, accuracy, trust, transparency, and general satisfaction. We also conduct a path analysis with different recommender attributes and user contextual factors. With that, we identify that low diversity and high serendipity in DL-RecSys models undermine transparency, trust, and accuracy of recommendations, ultimately compromising user satisfaction with the system. Finally, we analyze users' qualitative input, reveal their requests for beyond-accuracy recommendation attributes and different elements they use to build trust with the system. We hope this case study can serve as a new perspective on evaluating and optimizing future DL-RecSys models under a human-centric framework.

\bibliographystyle{ACM-Reference-Format}
\bibliography{sample-base}

\newpage
\appendix
\section{Appendices} \label{appendix}

\subsubsection*{Deep Learning Model Repositories:}

    \begin{itemize}

        \item \emph{\textbf{NCF}}: \url{https://github.com/yihong-chen/neural-collaborative-filtering}

        \item \emph{\textbf{SSE-PT}}: \url{https://github.com/lizli502/SSE-PT}

        \item \emph{\textbf{BERT4Rec}}: \url{https://github.com/jaywonchung/BERT4Rec-VAE-Pytorch}

        \item \emph{\textbf{GLocal-K}}: \url{https://github.com/fleanend/TorchGlocalK}

    \end{itemize}

    For all model reproduction, we mostly use the default optimized parameter claimed by the repository on the ml-1m dataset, detailed below. Specifically, we ran 301 epochs for each model and select the top 12 recommendations from the best performed eopch (based on its NDCG value) for each user.

\setcounter{table}{0}
\renewcommand{\thetable}{A\arabic{table}}

    \begin{table}[h!]
    \centering
    \resizebox{\columnwidth}{!}{%
    \begin{tabular}{@{}llll@{}}
    \toprule
    \textit{\textbf{}}          & \textbf{gmf\_config} & \textbf{mlp\_config} & \textbf{neumf\_config} \\ \midrule
    \textit{num\_epoch}         & 301                  & 301                  & 301                    \\
    \textit{batch\_size}        & 1024                 & 1024                 & 1024                   \\
    \textit{optimizer}          & adam                 & adam                 & adam                   \\
    \textit{adam\_lr}           & 1e-3                 & 1e-3                 & 1e-3                   \\
    \textit{latent\_dim}        & 8                    & 8                    & 8                      \\
    \textit{num\_negative}      & 4                    & 4                    & 4                      \\
    \textit{l2\_regularization} & 0.01                 & 0.0000001            & 0.01                   \\
    \textit{layers}             & -                    & {[}16,64,32,16,8{]}  & {[}16,64,32,16,8{]}    \\
    \textit{pretrain}           & -                    & True                 & True                   \\ \bottomrule
    \end{tabular}%
    }
    \caption{NCF training parameters}
    \label{tab:ncf_parameters}
    \end{table}

    \begin{table}[h!]
    \centering
    \resizebox{\columnwidth}{!}{%
    \begin{tabular}{llll}
    \hline
    \textbf{GLocal-K Param} & \textbf{GLocal-K Value} & \textbf{SSE-PT Param}        & \textbf{SSE-PT Value} \\ \hline
    \textit{n\_hid}         & 500                     & \textit{num\_epoch}          & 301                   \\
    \textit{n\_dim}         & 5                       & \textit{batch\_size}         & 128                   \\
    \textit{n\_layers}      & 2                       & \textit{max\_len}            & 50                    \\
    \textit{gk\_size}       & 3                       & \textit{user\_hidden\_units} & 50                    \\
    \textit{max\_epoch\_p}  & 500                     & \textit{item\_hidden\_units} & 50                    \\
    \textit{max\_epoch\_f}  & 500                     & \textit{lr}                  & 0.001                 \\
    \textit{patience\_p}    & 5                       & \textit{num\_blocks}         & 2                     \\
    \textit{patience\_f}    & 10                      & \textit{num\_heads}          & 1                     \\
    \textit{tol\_p}         & 1e-4                    & \textit{dropout\_rate}       & 0.5                   \\
    \textit{tol\_f}         & 1e-5                    & \textit{threshold\_user}     & 1.0                   \\
    \textit{lambda\_2}      & 20                      & \textit{threshold\_item}     & 1.0                   \\
    lambda\_s               & 0.006                   & l2\_emb                      & 0.0                   \\
    dot\_scale              & 1                       & k                            & 12                    \\ \hline
    \end{tabular}%
    }
    \caption{GLocal-K and SSE-PT training parameters}
    \label{tab:glocal-k-sse-pt_parameters}
    \end{table}

    \begin{table*}[]
    \centering
    \resizebox{1.5\columnwidth}{!}{%
    \begin{tabular}{@{}llllll@{}}
    \toprule
    \textit{\textbf{}}                      & \textbf{Dataset} & \textbf{Dataloader} & \textbf{NegativeSampler} & \textbf{Trainer} & \textbf{Model} \\ \midrule
    \textit{min\_rating}                    & 4                &                     &                          &                  &                \\
    \textit{min\_uc}                        & 5                &                     &                          &                  &                \\
    \textit{min\_sc}                        & 0                &                     &                          &                  &                \\
    \textit{split}                          & leave\_one\_out  &                     &                          &                  &                \\
    \textit{eval\_set\_size}                & 500              &                     &                          &                  &                \\
    \textit{train\_batch\_size}             &                  & 64                  &                          &                  &                \\
    \textit{test\_batch\_size}              &                  & 64                  &                          &                  &                \\
    \textit{train\_negative\_sampler\_code} &                  &                     & random                   &                  &                \\
    \textit{train\_negative\_sampler\_size} &                  &                     & 100                      &                  &                \\
    \textit{test\_negative\_sampler\_code}  &                  &                     & 100                      &                  &                \\
    \textit{test\_negative\_sampler\_size}  &                  &                     & 100                      &                  &                \\
    \textit{trainer\_code}                  &                  &                     &                          & bert             &                \\
    \textit{optimizer}                      &                  &                     &                          & adam             &                \\
    \textit{lr}                             &                  &                     &                          & 0.001            &                \\
    \textit{weight\_decay}                  &                  &                     &                          & 0                &                \\
    \textit{decay\_step}                    &                  &                     &                          & 15               &                \\
    \textit{gamma}                          &                  &                     &                          & 0.1              &                \\
    \textit{num\_epochs}                    &                  &                     &                          & 301              &                \\
    \textit{metric\_ks}                     &                  &                     &                          & {[}10,20,50{]}   &                \\
    \textit{best\_metric}                   &                  &                     &                          & NDCG@10          &                \\
    \textit{All BERT-relevant args}         &                  &                     &                          &                  & None           \\
    \textit{num\_item for DAE or VAE}       &                  &                     &                          &                  & None           \\
    \textit{num\_hidden for DAE or VAE}     &                  &                     &                          &                  & 0              \\
    \textit{hidden\_dim for DAE or VAE}     &                  &                     &                          &                  & 600            \\
    \textit{latent\_dim for DAE or VAE}     &                  &                     &                          &                  & 200            \\
    \textit{dropout for DAE or VAE}         &                  &                     &                          &                  & 0.5            \\ \bottomrule
    \end{tabular}%
    }
    \caption{BERT4Rec training parameters}
    \label{tab:bert4rec_parameters}
    \end{table*}
\end{document}